# Absence of Structural Impact of Noble Nanoparticles on P3HT: PCBM Blends for Plasmon Enhanced Bulk-Heterojunction Organic Solar Cells Probed by Synchrotron Grazing Incidence X-Ray Diffraction

Samuele Lilliu[1], Mejd Alsari[1], Oier Bikondoa[2], J. Emyr Macdonald[3], Marcus S. Dahlem[1]

[1]Masdar Institute of Science and Technology, PO Box 54224, Abu Dhabi, United Arab Emirates
[2]XMaS, The UK-CRG Beamline, ESRF-The European Synchrotron, CS40220, 38043 Grenoble Cedex 9, France
[3]School of Physics and Astronomy, Cardiff University, Queens Buildings, The Parade, Cardiff CF24 3AA, United Kingdom.

Correspondence and requests for materials should be addressed to Samuele Lilliu, samuele_lilliu@hotmail.it

The incorporation of noble metal nanoparticles, displaying localized surface plasmon resonance, in the active area of donor-acceptor bulk-heterojunction organic photovoltaic devices is an industrially compatible light trapping strategy, able to guarantee better absorption of the incident photons and give an efficiency improvement between 12% and 38%. In the present work, we investigate the effect of Au and Ag nanoparticles blended with P3HT: PCBM on the P3HT crystallization dynamics by synchrotron grazing incidence X-ray diffraction. We conclude that the presence of (1) 80nm Au, (2) mix of 5nm, 50nm, 80nm Au, (3) 40nm Ag, and (4) 10nm, 40nm, 60nm Ag colloidal nanoparticles, at different concentrations below 0.3 wt% in P3HT: PCBM blends, does not affect the behaviour of the blends themselves.

## Introduction

In the photovoltaics community it is commonly believed that in order to attract useful market shares, organic photovoltaic (OPV) solar cell efficiency must be higher than 10% for roll-to-roll-processed modules. In 2014, the best performing polymer lab-based triple junction devices achieved a power conversion efficiency (PCE) of ∼11.5%[1]. The photoactive area of the OPV devices discussed here is based on the bulk-heterojunction (BHJ) concept[2-4]. A BHJ is a thin film consisting of a blend between: (i) a p-doped electron-donor/hole-transporter phase (e.g. a polymer), and (ii) n-type, hole-donor/electron-transporter phase (e.g. a fullerene). Single junction OPVs typically consist of the following stacked layers: glass substrate, transparent conductive oxide, hole-donor interlayer (e.g. PEDOT: PSS), BHJ active layer (e.g. P3HT: PCBM), and top electrode (e.g. Ca/Al). In contrast to inorganic photovoltaic devices, OPVs are described as excitonic solar cells, to indicate that sufficiently energetic photons do not promptly produce free charges, but rather generate excitons[5]. Efficient charge harvesting from incoming photons imposes several morphological and optoelectrical constraints on the BHJ. The ideal BHJ nanostructure is a compromise between complete separation of the donor-acceptor phases and close interpenetration[5]. High recombination rate and low charge-carrier mobility limit the BHJ thickness to below 100-200 nm. This restricts the absorption yield and the resulting PCE[6]. Effective light trapping strategies offer a good approach to ensure better absorption of the incident photons, and a route for attaining efficiencies beyond the 10% barrier for single junction OPV, which would push the commercialization of organic solar cells[6].

Plasmonic nanostructures can be incorporated into the active area of organic solar cells to enhance the optical absorption and the current density, without increasing the thickness of their active areas[6]. Noble metal nanoparticles (NPs), such as Au and Ag, exhibit Localized Surface Plasmon Resonance (LSPR) that couple strongly to the incident light[7]. This effect can increase the light absorption capability of the OPV device within a broad range of wavelengths[7-10]. This strategy is highly compatible with industrial roll-to-roll OPV fabrication processes, since the NPs are easily blended within the BHJ solution. To date, the most relevant research in this field has been conducted by D. H. Wang et al. since 2011[8,9,11]. The authors demonstrate that plasmon enhanced bulk-heterojunction OPVs show a PCE increase of 12-38%, when compared to control devices without NPs. Still missing in the literature is a comprehensive structural investigation of NPs enhanced OPVs, and a discussion on the correlation between structural properties of the films and device properties. A Grazing Incidence Wide Angle X-Ray Scattering (GI-WAXS) study recently showed that, by introducing $Cu_2S$ NPs in the BHJ, the self-organized nano-structural evolution of the donor-acceptor phases can be finely tuned[12]. The main question is whether a similar effect can be observed with noble metal NPs blended with P3HT: PCBM. The investigation of donor-acceptor Grazing Incidence X-Ray Diffraction (GI-XRD) patterns before and after the annealing is a powerful tool for understanding the evolution dynamics of the nano-morphology of the two materials during thermal or solvent annealing[5,13-19]. Information extracted from GI-XRD patterns of NPs enhanced BHJ could be correlated to other studies to improve understanding of the morphological factors affecting the overall OPV performance[16].

To the best of our knowledge, there are not any reported comprehensive synchrotron GI-XRD investigations on the effects of Au and Ag NPs on the structural properties of the donor-acceptor phases in OPVs bulk-heterojunctions. In this work we investigate the effect of the presence of Au and Ag NPs in the P3HT: PCBM bulk-heterojunction on the semi-crystalline P3HT phase.

## Experimental Details

Colloidal Au nanoparticles suspended in a citrate buffer solution were purchased from Sigma-Aldrich (5nm, 741949-25ML, 50nm, 742007-25ML, 80nm, 742023-25ML). Colloidal Ag nanoparticles dispersed in aqueous buffer containing sodium citrate as the stabilizer were purchased from Sigma-Aldrich (10nm, 730785-25ML, 40nm, 730807-25ML, 60nm, 730815-25ML). Regioregular poly(3-hexylthiophene-2,5-diyl) (P3HT) was purchased from Ossila Ltd (M107, RR 93.6%, Mw 31,300, Mn 15,600). A mixture of [6,6]-Phenyl-C71-butyric acid methyl ester (PC70BM) and [6,6]-Phenyl-C61-butyric acid methyl ester (PC60BM) (95/5%) was purchased from Ossila Ltd (M113). Silicon oxide ($Si/SiO_2$) substrates were purchased from Ossila Ltd, and cleaved into substrates smaller than $1cm^2$. The substrates were cleaned by sonication in water with Hellmanex (Z805939, FLUKA), deionized water, acetone, isopropanol, and water (15min each). A P3HT: PCBM solution was prepared by dispersing 25mg/mL of P3HT and PCBM (1:0.6) in chlorobenzene. The solution was stirred at 80°C for 10min and overnight at 50°C. Twelve 4mL vials containing different amounts of nanoparticles were left on a hotplate at 100°C for several hours until the solvent was completely evaporated[20]. 33mL of P3HT: PCBM blend was then added to each vial. The final solution was stirred overnight at 50°C. The list of solutions and P3HT: PCBM: NPs wt% is reported in Table 1. 'Au mix' refers to a mix between 5nm (33%), 50nm (33%) and 80nm (33%) Au nanoparticles. 'Ag mix' refers to a mix between 10nm (33%), 40nm (33%) and 60nm (33%) Ag nanoparticles. Each solution was spin-coated on the clean silicon oxide substrates at 2000rpm for 30min under nitrogen atmosphere. Half of the samples were ex-situ annealed before the XRD measurements for 15min at 150°C.

Table 1 – P3HT: PCBM: Au/Ag NPs solutions, amount of NPs from the as-received solution, and P3HT: PCBM: NPs wt%.

| Solution number | NPs | Amount | P3HT % | PCBM % | NPs % |
|---|---|---|---|---|---|
| 1 | Au 80nm | 1mL | 62.37 | 37.42 | 0.20 |
| 2 | Au 80nm | 0.5mL | 62.44 | 37.46 | 0.10 |
| 3 | Au 80nm | 0.25mL | 62.47 | 37.48 | 0.05 |
| 4 | Au mix | 1mL | 62.33 | 37.40 | 0.28 |
| 5 | Au mix | 0.5mL | 62.41 | 37.45 | 0.14 |
| 6 | Au mix | 0.25mL | 62.48 | 37.49 | 0.03 |
| 7 | Ag 40nm | 1mL | 62.45 | 37.47 | 0.08 |
| 8 | Ag 40nm | 0.5mL | 62.48 | 37.49 | 0.04 |
| 9 | Ag 40nm | 0.25mL | 62.49 | 37.49 | 0.02 |
| 10 | Ag mix | 1mL | 62.45 | 37.47 | 0.08 |
| 11 | Ag mix | 0.5mL | 62.48 | 37.49 | 0.04 |
| 12 | Ag mix | 0.25mL | 62.49 | 37.49 | 0.02 |

The average sample thickness was 100nm. Details of our GI-XRD setup and geometry have already been reported[13-16,21,22]. GI-XRD measurements were performed at the XMaS beamline (BM28, ESRF, Grenoble, France). Diffraction images were collected with a 2D detector (MAR SX-165). The X-ray energy was 10keV. The critical angle of the films was $α_c ≈ 0.12°$, and measurements were taken at different incident angles $α_i$ below and above the critical angle. The Matlab software developed by Dr. Lilliu for the GI-XRD analysis[14] was updated with a new routine for total image remapping into the scattering vector ($q$) space. The $q_{xy} = 0$, $q_z = 1$ [Å$^{-1}$] coordinate on the diffraction images corresponds to the azimuthal angle $χ = 0°$ (out-of-plane direction, or OOP), while the $q_{xy} = 1$, $q_z = 0$ [Å$^{-1}$] direction, corresponds to $χ = 90°$ (in-plane direction, or IP). Therefore, due to the experimental geometry, the cake slice corresponding to $χ = 0°$ is not accessible in the reported diffraction images (black background). Line profiles are extracted from cake slices at different azimuthal angles with an integration aperture of $Δχ = 10°$. Peak fitting is performed as previously reported[14].

## Results and Discussion

Following the study shown by Liao et al. on the effect of $Cu_2S$ NPs on the nano-structural evolution of the donor-acceptor phases[12], we investigate the effect of commercial Au and Ag NPs in the P3HT: PCBM bulk-heterojunction. The colloidal NPs employed here, have an absorption spectrum[23] that overlaps with the visible spectrum and can be employed in the fabrication of OPVs based on plasmon enhanced BHJ. NPs have a bell-like absorption spectrum with a peak in the visible that varies with the size of the NPs[23]. Mixing NPs with different sizes allows achieving a broad absorption spectrum and the possibility to exploit the plasmonic effect on a broader range of wavelengths.

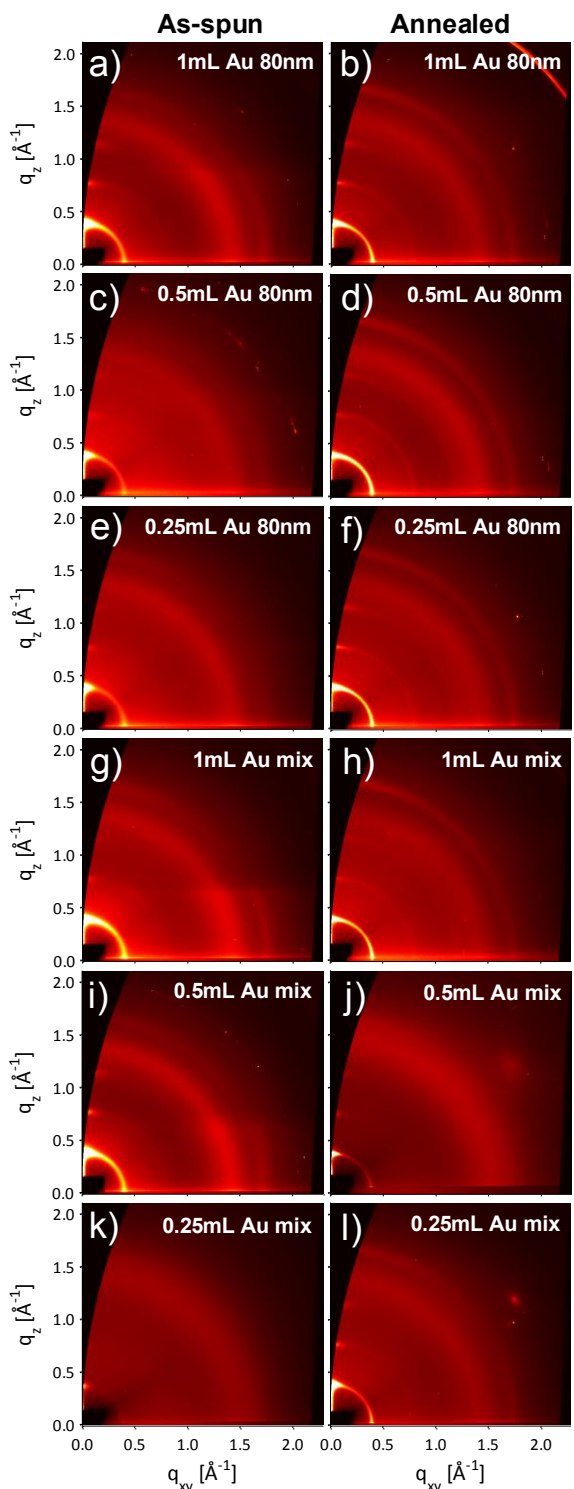

**Figure 1 – Diffraction images collected at the critical angle for as-spun and annealed P3HT: PCBM: Au NPs blends: (a-b) Au 80nm 1mL, (c-d) Au 80nm 0.5mL, (e-f) Au 80nm 0.25mL, (g-h) Au mix 1mL, (i-j) Au mix 0.5mL, and (k-l) Au mix 0.25mL.**

Figure 1 shows diffraction patterns collected at $\alpha_i = 0.12°$ for P3HT: PCBM: Au NPs films before and after ex-situ annealing. Images (a-f) are from blends mixed with 80nm Au NPs (solutions 1-3 in Table 1), while images (g-l) are from blends with a mix of 5nm, 50nm and 80nm (solutions 4-6 in Table 1). Figure 2 shows line profiles extracted along the OOP direction ($\chi = 15°$), for the blends with (a) 80nm Au NPs, and (b) Au mix.

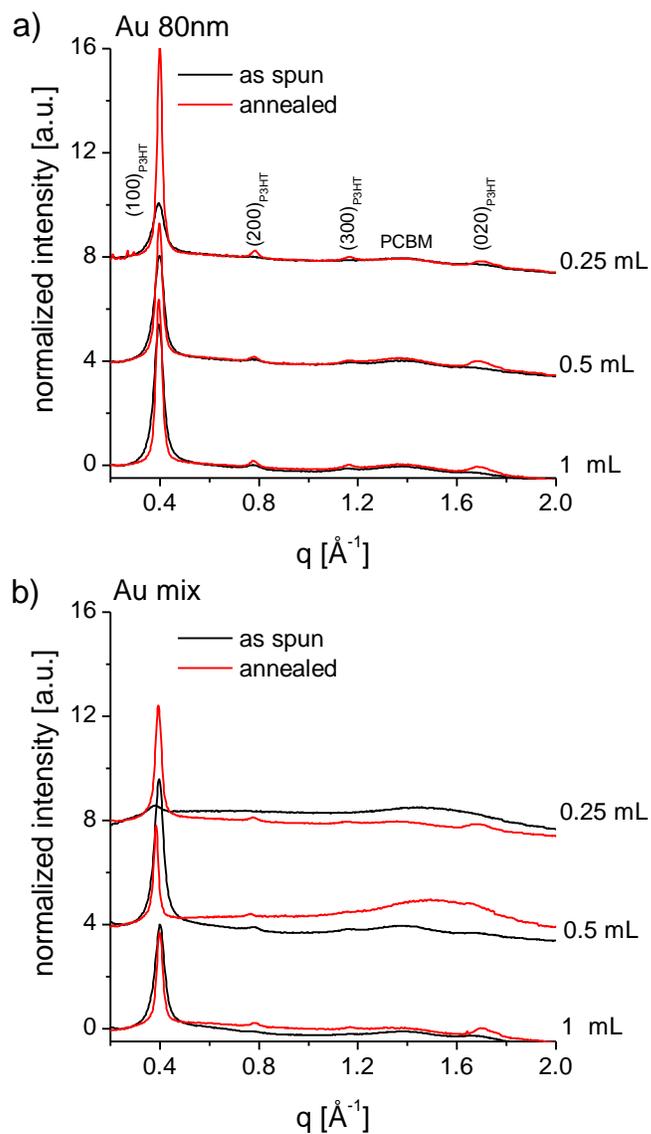

**Figure 2 – Line profiles collected at critical angle, extracted with cake slices centred at $\chi = 15°$ ($\Delta\chi = 10°$ integration aperture), for films deposited from P3HT: PCBM: Au NPs solutions (1-6), measured ex-situ before and after annealing. All line profiles are normalized with respect to the line profile region between the end of the beamstop and the beginning of the first peak, the (100)-P3HT peak. P3HT and PCBM peaks are indicated in (a).**

The diffraction patterns resemble those reported in our previous work, and diffraction peaks have been indexed accordingly[13-15]. P3HT crystallizes with a monoclinic unit cell[14,24], and self-organizes into lamellar structures[25-28]. In the lamella, the backbones are oriented along the *c*-direction. Continuous lamellae are stacked (i) orthogonal to the parallel conjugated backbones in the *b*-direction and (ii) along the alkyl-stacking *a*-direction[14]. The alkyl-stacking direction can be perpendicular (edge-on lamellae) to the sample substrate or parallel to the sample substrate (face-on lamellae)[14]. PCBM crystallizes from chlorobenzene with a triclinic unit cell, and does not preferentially orientate with respect to the sample substrate, under usual processing conditions[14,29-31].

In order to verify whether the presence of NPs in the blend induces structural changes in the semi-crystalline phase of P3HT, line profiles extracted at $\chi = 15°$ for angles above the critical angle are fitted as previously described[13-15,21]. At these angles, the retrieved information comes both from the bulk and the surface. The P3HT domain size along the *a*-direction of the alkyl side chains is estimated from the full width at half maximum (FWHM) of the (100)-P3HT peaks. The estimated domain sizes are calculated using the Scherrer equation[32], i.e. $L_{100} \approx 0.9 \times 2\pi / \text{FWHM}$, which usually overestimates the real domain size and does not take into account disorder contributions in the peak broadening[14]. The purpose of this initial analysis is to verify whether there are significant differences in $L_{100}$ among different samples. Figure 3 shows the estimated domain sizes $L_{100}$ for the 6 spin-coated solutions. It is clear that different concentrations of NPs do not affect the domain size, and that the difference in $L_{100}$ between samples is simply due to stochastic effects. A common aspect to all the samples is the well-known increase in the P3HT domain size due to the crystallization induced by the anneal[14].

Given the result, further investigation, such as the Williamson-Hall or the Warren-Averbach[32] analysis for the separation of the correlation length contribution from the disorder contribution in the peak broadening, would not add any significant information.

Figure 4 shows the estimated P3HT lattice constant along the *a*-direction of the alkyl-stacking chains. Refraction effects have not been taken into account. The (100)-P3HT peak plots vs $\alpha_i$ (not shown) all show the characteristic profile, with a peak located at the critical angle $\alpha_c = 0.12°$, with the exception of Au mix 0.5mL, which was misaligned. Again, we do not observe any trend as a function of different concentrations of Au NPs. A common characteristic observed in the literature is the expansion of the a lattice constant after the anneal[14]. Interestingly, this condition is not verified for both samples containing 0.25mL of NPs.

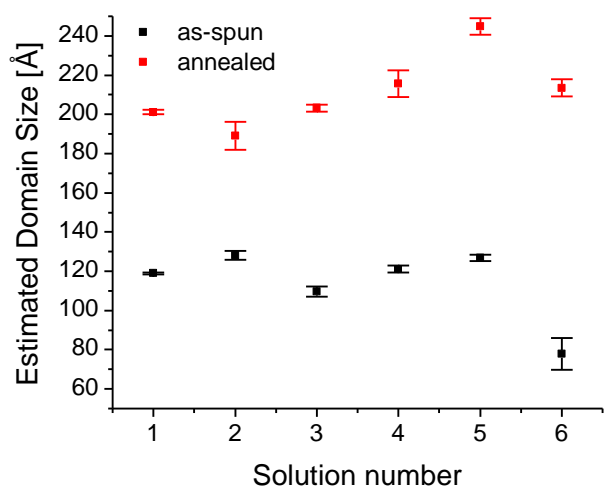

**Figure 3** – (100)-P3HT domain sizes estimated with the Scherrer equation for P3HT: PCBM: Au NPs. Statistics are performed on several images collected at different angles above $\alpha_i = 0.16°$, in particular: (i) as-spun samples, 10 for solution 1 (80nm, 1mL), 3 for solution 2 (80nm, 0.5mL), 9 for solution 3 (80nm, 0.25mL), 10 for solution 4 (mix, 1mL), 4 for solution 5 (mix, 0.5mL), and 14 for solution 6 (mix, 0.25mL); (ii) annealed samples, 5 for solution 1 (80nm, 1mL), 7 for solution 2 (80nm, 0.5mL), 5 for solution 3 (80nm, 0.25mL), 6 for solution 4 (mix, 1mL), 15 for solution 5 (mix, 0.5mL), and 13 for solution 6 (mix, 0.25mL).

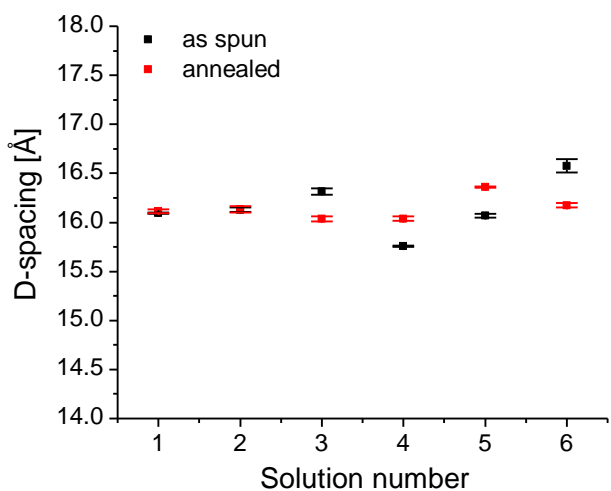

**Figure 4** – (100)-P3HT lattice constant along the *a*-direction of the alkyl-stacking chains for P3HT: PCBM: Au NPs. See Figure 3 for information on the statistics.

The same type of analysis is performed for the samples containing Ag. Figure 5 shows diffraction patterns collected at the critical angle for P3HT: PCBM: Ag NPs films before and after ex-situ annealing. Images (a-f) are from blends mixed with 40nm Ag NPs (solutions 7-9 in Table 1), while images (g-l) are from blends with a mix of 10nm, 40nm and 60nm (solutions 10-12 in Table 1). Figure 6 shows line profiles extracted along the OOP direction ($\chi = 15°$), for the blends with (a) 40nm Ag NPs, and (b) Ag mix.

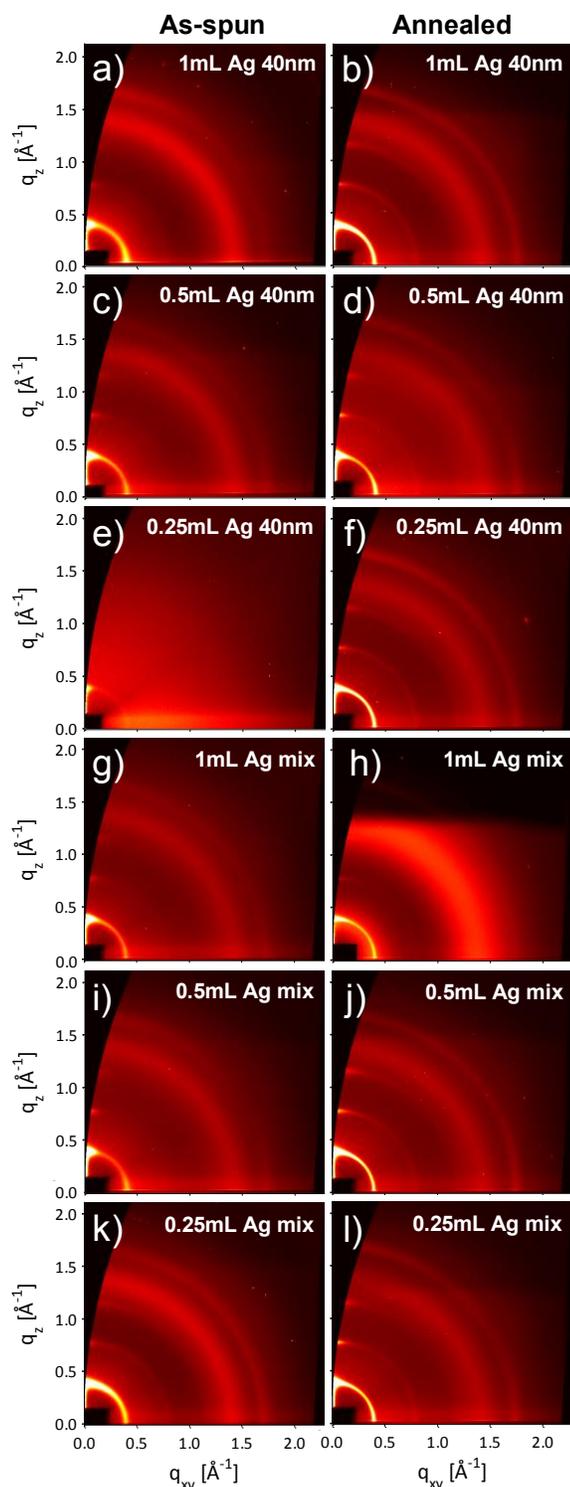
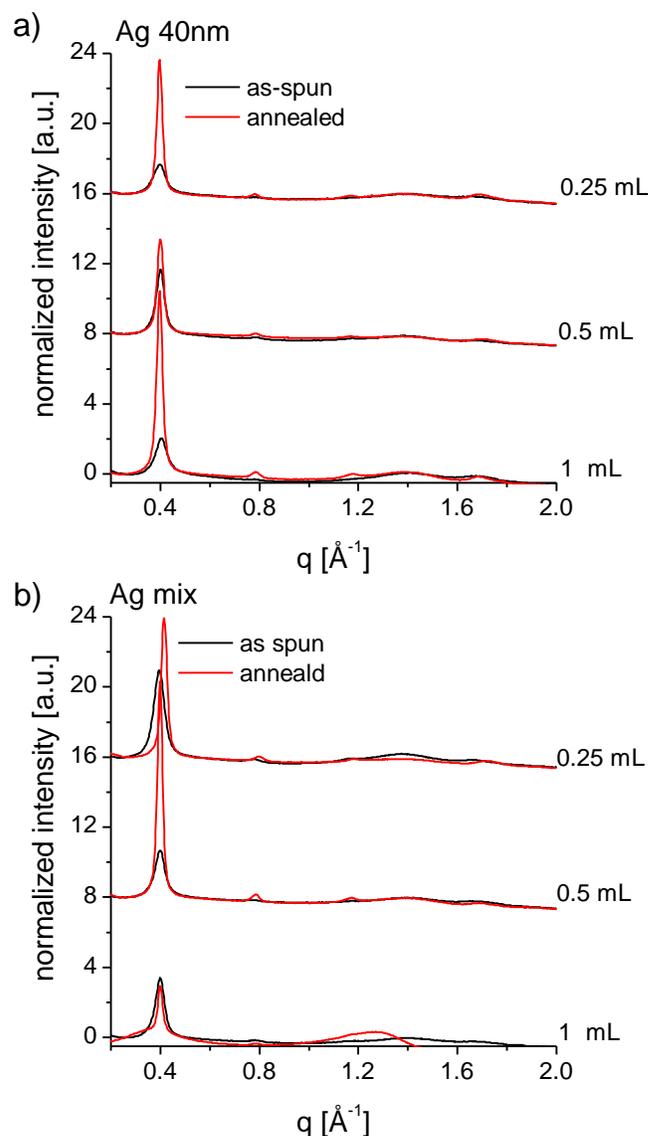

**Figure 5** – Diffraction images collected at the critical angle for as-spun and annealed P3HT: PCBM: Ag NPs blends: (a-b) Ag 40nm 1mL, (c-d) Ag 40nm 0.5mL, (e-f) Ag 40nm 0.25mL, (g-h) Ag mix 1mL, (i-j) Ag mix 0.5mL, and (k-l) Ag mix 0.25mL. Images are extracted at the critical angle in order to display the maximum intensity.

**Figure 6** – Line profiles collected at critical angle, extracted with cake slices centred at $\chi = 15°$ ($\Delta\chi = 10°$ integration aperture), for films deposited from P3HT: PCBM: Ag NPs solutions (7-12), measured ex-situ before and after annealing. All line profiles are normalized with respect to the line profile region between the end of the beamstop and the beginning of the first peak, (100)-P3HT peak. The line profile drop near $q = 1.5$Å$^{-1}$ for the annealed 1mL in (b) is due to background issues.

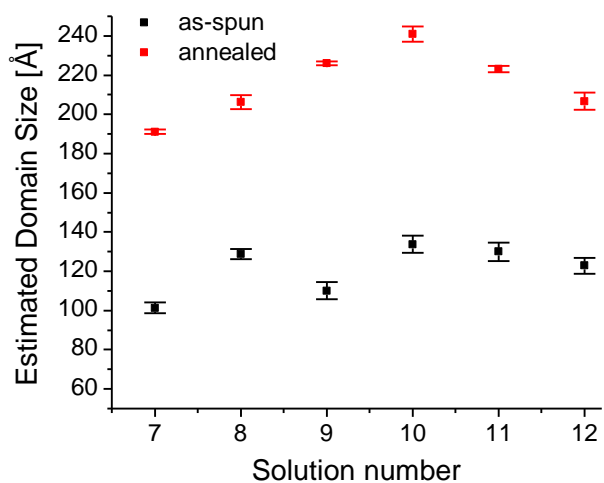

**Figure 7** – (100)-P3HT domain sizes estimated with the Scherrer equation for P3HT: PCBM: Ag NPs. Statistics are performed on several images collected at different angles above $\alpha_i = 0.16°$, in particular: (i) as-spun samples, 11 for solution 7 (40nm, 1mL), 13 for solution 8 (40nm, 0.5mL), 8 for solution 9 (40nm, 0.25mL), 6 for solution 10 (mix, 1mL), 12 for solution 11 (mix, 0.5mL), and 8 for solution 12 (mix, 0.25mL); (ii) annealed samples, 10 for solution 7 (40nm, 1mL), 12 for solution 8 (40nm, 0.5mL), 10 for solution 9 (40nm, 0.25mL), 11 for solution 10 (mix, 1mL), 11 for solution 11 (mix, 0.5mL), and 11 for solution 12 (mix, 0.25mL).

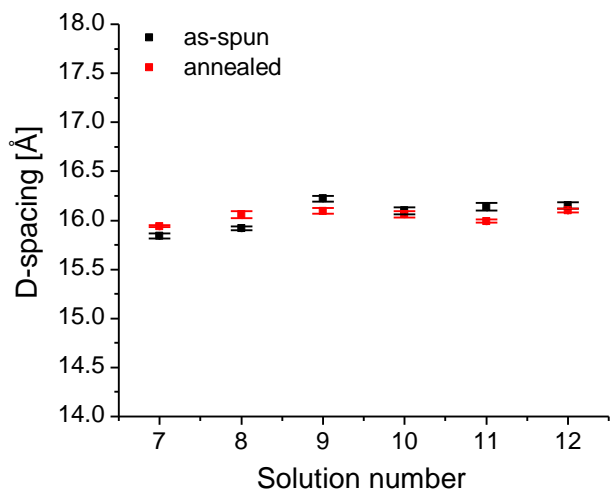

**Figure 8** – (100)-P3HT lattice constant along the a-direction of the alkyl-stacking chains for P3HT: PCBM: Ag NPs. See Figure 7 for information on the statistics.

Figure 7 shows the estimated (100)-P3HT domain size for the samples with Ag NPs. The trend observed for the annealed samples from solutions with 40nm Ag NPs (7-9) is opposite to the trend observed for the samples prepared with the mix solutions (10-12). Apart from the as-spun sample from solution 8, as-spun samples follow the same trend of the annealed samples. The estimated domain sizes $L_{100}$ are within the same range of values shown for the samples containing Au NPs. Similarly to the samples containing Au NPs, we believe the difference in (100)-P3HT domain sizes between samples is simply due to stochastic effects. Figure 8 shows the estimated lattice constant along the *a*-direction. Also in this case, we do not observe any trend as a function of the concentration of NPs. A qualitative comparison between the estimated lattice constants in Figures 4 and 8 indicates that the graphs are not correlated.

## Conclusion

The experiments presented in this work were carried out to investigate the potential effect of noble nanoparticles on the structural properties of the semi-crystalline P3HT phase in P3HT: PCBM blends, used in plasmon enhanced organic bulk-heterojunction solar cells. We conclude that there is no observable correlation between the concentration of nanoparticles dispersed in the P3HT: PCBM blend and the estimated P3HT domain size and lattice constant along the alkyl-stacking direction. This could be due to the fact that the colloidal nanoparticles do not interfere with the bulk-heterojunction crystallization, or to the fact that the interference is not measurable at low (below 0.3 wt%) nanoparticle concentrations.

## Acknowledgements

We would like to thank the BM28 (XMaS, ESRF, Grenoble, France) team for help at the beamline, and R. Tucker (Cardiff University) for help with the instrumentation. Financial support was provided by the University of Warwick through the Engineering and Physical Sciences Research Council, and by Masdar Institute. We would like to thank B. Curzadd, M. A. Davis, and R. Song (New York University) for manufacturing the XRD chamber, and C. Maragliano for help with the measurements.